\documentclass{epl2}
\usepackage{amsmath}

\newcommand{\rcite}[1]{Ref.~\citenum{#1}}
\newcommand{\eq}[1]{Eq.~(\ref{#1})}
\newcommand{\ez}{{\bm e}_z}
\newcommand{\bn}{{\bm n}}

\title{ {\large 
Comment on ``Capillary attraction of charged particles at a
  curved liquid interface'' by A. W{\"u}rger}\\[3mm]
Absence of logarithmic attraction between colloids trapped at the interface of droplets} \shorttitle{Comment to EpL {\bf 75} (2006) 978}

\author{A. Dom{\'\i}nguez\inst{1} \and M. Oettel\inst{2} \and S. Dietrich\inst{3}}
\shortauthor{Dom{\'\i}nguez, Oettel, Dietrich}
\institute{
  \inst{1} F\'\i sica Te\'orica, U.\ Sevilla, Apdo.~1065, E--41080 Sevilla, Spain (email: {\tt dominguez@us.es})\\
  \inst{2} 
  Institut f\"ur Physik, WA331, U.\ Mainz, D-55099 Mainz, Germany \\
  \inst{3} 
  MPI f\"ur Metallforschung, Heisenbergstr.~3, D--70569 Stuttgart, Germany and \\
  ITAP, U.\ Stuttgart, Pfaffenwaldring 57, D--70569 
  Stuttgart, Germany
}


\begin{document}

\maketitle


The possibility of long--ranged attractions between colloids
trapped at fluid interfaces is a topic of current interest.
\rcite{Wuer06b} proposes an intriguing mechanism of geometrical nature 
for the appearance of a logarithmic capillary attraction between colloids on a 
droplet. In this respect
we note that such an attraction was already proposed
in \rcite{ODD05a} with essentially the same physical interpretation 
as the one given in \rcite{Wuer06b}, i.e., an 
unbalance between the forces on the particle and on the curved interface, 
respectively.
However, following a different approach it was shown that such an
unbalance does not arise \cite{DOD05}. Without being exhaustive, here we correct only
those mistakes in \rcite{Wuer06b} which affect the main conclusion. We 
demonstrate that there is no logarithmically varying
interfacial deformation and consequently no logarithmic capillary attraction. 
Incidentally, this Comment provides an
independent confirmation of the conclusions obtained in \rcite{DOD05}\footnote{Roman equation numbers will
  refer to this Comment and arabic equation numbers to
  \rcite{Wuer06b}.}.

\medskip\noindent (a) {\it The definition of $\Pi_1$}: For an isolated
droplet in equilibrium, the electrical force ${\bm K}=K \bn_Q$ acting
on the particle is compensated by tension exerted by the interface at
the contact line $\theta=\theta_0$. In \rcite{Wuer06b}, the effect of
this line tension is modelled by a pressure term $\Pi_1$ acting along
the normal $\bn_0$ of the undeformed interface.
The value $\Pi_1$ of this model must be determined by Eq.~(9), i.e.,
Newton's law of action--reaction.
Exploiting rotational symmetry, Eq.~(9) can be projected onto the unit vector $\ez(=\bn_Q)$
and evaluated with Eq.~(11):
\begin{equation}
  \label{eq:mechisol}
  0 = \int \upd\Omega \, (\bn\cdot\ez) P(\bn) = 
  2\pi \int_{\theta_0}^\pi \upd\theta \, \sin\theta \cos\theta \, 
  \Pi(\theta) - \Pi_1 \cos\theta_0 .
\end{equation}
This corrects the definition of $\Pi_1$ in Eq.~(10) by a factor
$1/\cos^2\theta_0$. The origin for this discrepancy is that the
force--balance condition~(9) actually means that the projection of
$\Pi_1$ onto $\bn_Q(=\ez)$ must equal ${\bm K}$, whereas
\rcite{Wuer06b} imposes that the projection of ${\bm K}$ onto $\bn_0$
must equal $\Pi_1$, which violates the action--reaction principle.

\medskip\noindent
(b) {\it A useful intermediate result}: Given the rotational
symmetry, Eq.~(5) can be rewritten as
\begin{equation}
  \label{eq:rev_u} u(\theta) = \frac{R^2}{\gamma} \int_0^\pi
  \upd\theta' \, \sin\theta' g(\theta,\theta') P(\theta') , \qquad
  \textrm{with}\quad g(\theta,\theta') := \int_0^{2\pi} \upd\varphi' \,
  G(\bn,\bn') .
\end{equation} 
For $\theta'\leq\theta\leq\pi$ the Green function $g(\theta,\theta')$ is given
by
\begin{equation}
  \label{eq:gup} g(\theta,\theta') = \frac{1}{2}\cos\theta [1 -
  \cos\theta' \ln (1+\cos\theta')] - \frac{1}{2}\cos\theta' [1 +
  \cos\theta \ln (1-\cos\theta)] + \frac{\ln 2-1}{2}\cos\theta
  \cos\theta' ,
\end{equation} while for $0\leq\theta\leq\theta'$ it has this same form but with $\theta$ and $\theta'$ exchanged.

\medskip\noindent
(c) {\it The solution $u_L(\theta)$} is given by \eq{eq:rev_u}
with $P_L(\theta)$ defined in \rcite{Wuer06b} above Eq.~(12):
\begin{equation}
  \label{eq:uL}
  u_L(\theta) = \frac{R^2 (\Pi_0-\Pi_1)}{2\pi\gamma} g(\theta,\theta_0) 
  = - \frac{R^2 (\Pi_0-\Pi_1)}{4\pi\gamma} \cos\theta_0 
  [\cos\theta \ln(1-\cos\theta) + 1 ] + A\cos\theta , 
\end{equation}
where $\theta_0<\theta$ and the constant $A$ depends on $\theta_0$.
This expression corrects Eq.~(12) in two aspects: (i) the presence of
a term $\propto \cos\theta$, which however does not affect the
amplitude of a possible logarithm in the range
$\theta_0<\theta\ll\pi$, and (ii) the amplitude of the logarithm,
which has an additional factor $\cos\theta_0$ as well as the opposite
sign as the one given in \rcite{Wuer06b}.\footnote{The sign can be guessed already from the comparison with
  the flat--interface limit of the solution ($\theta_0,\theta\to 0$):
  if $\Pi_0-\Pi_1 >0$ (interface is pulled up), it follows that
  $u'(\theta)<0$.}.

\medskip\noindent
(d) {\it The solution $u_A(\theta)$:} Applying again \eq{eq:rev_u}
with the stress $P-P_L$, we find
\begin{equation}
  \label{eq:uA}
  u_A(\theta) = \frac{R^2}{\gamma} \int_{\theta_0}^\pi \upd\theta' \, \sin\theta' 
  g(\theta,\theta') \Pi(\theta') - \frac{R^2 \Pi_0}{2\pi\gamma} g(\theta,\theta_0) .
\end{equation}
In the integral term, a logarithmic dependence is given by
\begin{eqnarray}
  \label{eq:asymp}
  \frac{R^2}{\gamma} \int_{\theta_0}^\pi \upd\theta' \, \sin\theta'
  g(\theta,\theta') \Pi(\theta') & = &
  -\frac{1}{2} \cos\theta \ln(1-\cos\theta)
  \int_{\theta_0}^\pi \upd\theta' \, \sin\theta' \cos\theta' \Pi(\theta') +\dots
\end{eqnarray}
All other contributions are either manifestly unimportant or 
do not yield logarithmic terms if $\Pi(\theta)$ decays rapidly enough. 
To leading order in
$\theta_0\ll\pi$, the amplitude of the logarithm is canceled by the
term $\propto \Pi_0$ in \eq{eq:uA}, but there still remains a
logarithm with an amplitude ${\cal O}(\theta_0^2)$. The key point is
that the amplitude of the logarithm in \eq{eq:uL} is also ${\cal
  O}(\theta_0^2)$, so that the solution $u_A(\theta)$ has to be
expanded up to and including ${\cal O}(\theta_0^2)$ in order to compare it
with $u_L(\theta)$ consistently. Therefore Eq.~(17) is missing a
logarithmic term with an amplitude which is comparable to that of the logarithm 
in Eq.~(12).

\medskip\noindent (e) 
{\it The total deformation} is given by 
Eqs.~(\ref{eq:uL}), (\ref{eq:uA}), and (\ref{eq:asymp}):
\begin{equation}
  u = u_L + u_A = \frac{R^2}{4\pi\gamma} \cos\theta \ln(1-\cos\theta) 
  \left[ \Pi_1 \cos\theta_0 - 
    2\pi \int_{\theta_0}^\pi \upd\theta' \, \sin\theta' \cos\theta' \Pi(\theta') 
  \right] + \dots 
\end{equation}
ignoring terms which do not contribute to a logarithmic dependence for
$\theta\ll\pi$. From \eq{eq:mechisol} we conclude that the amplitude of the
logarithm vanishes. This is a general, exact result which
holds to all orders in $\theta_0$, in agreement with \rcite{DOD05}.
  

\end{document}